\documentclass[aip,amsmath,amssymb,reprint]{revtex4-1}
\usepackage{graphicx}
\usepackage{dcolumn}
\usepackage{bm}
\usepackage[utf8]{inputenc} 
\usepackage[T1]{fontenc}
\usepackage{mathptmx}
\usepackage{gensymb}
\usepackage{ragged2e}

\begin{document}
\title{Cavity-Enhanced 2D Material Quantum Emitters Deterministically Integrated with Silicon Nitride Microresonators}

\author{K. Parto}
\altaffiliation{These authors contributed equally to this work.}
\affiliation{Electrical and Computer Engineering Department, University of California, Santa Barbara, CA 93106, USA}

\author{S. I. Azzam}
\altaffiliation{These authors contributed equally to this work.}
\affiliation{Electrical and Computer Engineering Department, University of California, Santa Barbara, CA 93106, USA}
\affiliation{California Nanosystems Institute, University of California, Santa Barbara, CA 93106}
\author{N. Lewis}
\author{S. D. Patel}
\author{S. Umezawa}
\affiliation{Electrical and Computer Engineering Department, University of California, Santa Barbara, CA 93106, USA}
\author{K. Watanabe}
\affiliation{Research Center for Functional Materials, National Institute for Materials Science, 1-1 Namiki, Tsukuba 305-0044, Japan}
\author{T. Taniguchi}
\affiliation{International Center for Materials Nanoarchitectures, National Institute for Materials Science, 1-1 Namiki, Tsukuba 305-0044, Japan}
\author{G. Moody}
\email{moody@ucsb.edu}
\affiliation{Electrical and Computer Engineering Department, University of California, Santa Barbara, CA 93106, USA}
\affiliation{California Nanosystems Institute, University of California, Santa Barbara, CA 93106}

\date{\today}

\begin{abstract}
Optically active defects in 2D materials, such as hexagonal boron nitride (hBN) and transition metal dichalcogenides (TMDs), are an attractive class of single-photon emitters with high brightness, room-temperature operation, site-specific engineering of emitter arrays, and tunability with external strain and electric fields. In this work, we demonstrate a novel approach to precisely align and embed hBN and TMDs within background-free silicon nitride microring resonators. Through the Purcell effect, high-purity hBN emitters exhibit a cavity-enhanced spectral coupling efficiency up to $46\%$ at room temperature, which exceeds the theoretical limit for cavity-free waveguide-emitter coupling and previous demonstrations by nearly an order-of-magnitude. The devices are fabricated with a CMOS-compatible process and exhibit no degradation of the 2D material optical properties, robustness to thermal annealing, and 100 nm positioning accuracy of quantum emitters within single-mode waveguides, opening a path for scalable quantum photonic chips with on-demand single-photon sources.
\end{abstract}

\maketitle
\section*{\label{sec:intro}Introduction}
\vspace{-5pt}


\noindent Solid-state single-quantum emitters (SQEs) integrated with chip-scale photonic circuitry are a key building block for quantum information technologies \cite{Lodahl2015,Senellart2017,Moody_2022}. Numerous SQEs capable of high-purity emission at room-temperature have been discovered in several host materials, including diamond\cite{Aharonovich_2011}, silicon nitride\cite{senichev2021room}, and hexagonal boron nitride (hBN)\cite{tran2016quantum}. The integration of SQEs with CMOS-compatible photonic platforms would address a long-standing need for combining the manufacturability and scalability inherent to silicon-based photonics with materials that host high-quality SQEs. Heterogeneous integration techniques have led to successful demonstrations at cryogenic temperatures, including arrays of SQEs in diamond coupled to aluminum nitride photonic integrated circuits (PICs)\cite{Wan2020} and self-assembled quantum dots integrated with silicon nitride\cite{davanco2017heterogeneous,Chanana2022}. Yet, scalable strategies for the integration of SQEs operating at room temperature with silicon-based PICs have not yet been demonstrated. Four key requirements are necessary to address this challenge: (1) A host material with high-purity and bright emitters at room temperature; (2) the ability to reliably integrate the SQE host material with the PIC platform without altering the optical properties; (3) control of the emission wavelength and precise alignment within low-loss and background-free single-mode waveguide structures; and (4) integration with microresonators for efficient emission into a single optical mode through the Purcell effect.

\begin{figure*}[t] \centering
     \includegraphics[scale=1.1]{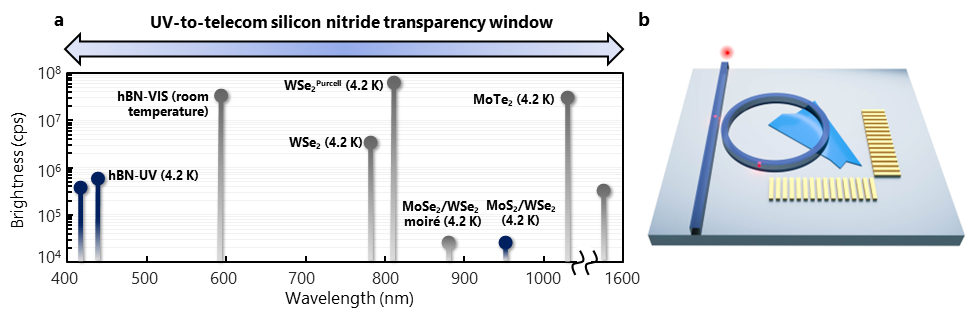}  
          \caption{\textbf{A universal platform for precision integration of 2D quantum emitters in silicon nitride photonics.} \textbf{a}, The family of 2D materials, including hBN emitters at UV \cite{fournier2020position,gale2022site} and visible \cite{grosso2017tunable} wavelengths, TMD monolayers \cite{luo2018deterministic,peyskens2019integration,zhao2021site}, and TMD heterostructures\cite{zhao2022manipulating,baek2020highly}, exhibits a rich spectrum of quantum emitters spanning the ultraviolet-to-telecommunications wavelength transparency window of silicon nitride photonics. The height of each bar indicates the reported intensity of the photoluminescence from the class of emitters (the gray data points are brightness-corrected for the objective extraction efficiency, whereas the blue data points are reported at the detector). \textbf{b}, Concept for the deterministic integration of a 2D material quantum emitter embedded within a silicon nitride microresonator with its electric dipole aligned to maximize overlap with the cavity modes for large Purcell enhancement and coupling efficiency.}  
          \label{fig:intro}
\end{figure*}

Of the SQE platforms, defect-type emitters in 2D materials \cite{toth2019single,kianinia2022quantum,turunen2022quantum,Azzam2021} have emerged as an attractive approach for engineering high-quality single-photon sources. SQEs have been identified in several monolayer and few-layer 2D materials spanning ultra-violet to telecommunications wavelengths, including hBN\cite{tran2016quantum,tran2016robust,fournier2020position}, transition metal dichalcogenides (TMDs) \cite{koperski2015single,srivastava2015optically,tonndorf2015single, he2015single, chakraborty2015voltage,Branny2016MoSe2,chakraborty2016localized,zhao2021site}, and heterostructures \cite{baek2020highly,zhao2021site}. In hBN and WSe$_2$, above $10$ MHz emission rates \cite{grosso2017tunable,luo2018deterministic} and 95$\%$ single-photon purity have been reported. Through strain and defect engineering, emitters can be aligned into arrays \cite{palacios2017large,parto2021defect}, and nanophotonic integration further enhances their brightness \cite{luo2019single,sortino2021bright}. SQEs in hBN are particularly appealing due to a 5.7 eV bandgap, which enables room temperature generation of single photons with up to 93$\%$ purity \cite{grosso2017tunable}. 

Thin layers of 2D materials can be aligned and placed onto a variety of substrates with automated dry-transfer techniques. These methods have enabled the integration of WSe$_2$ and hBN with silicon nitride\cite{peyskens2019integration,errando2021resonance,elshaari2021deterministic}, which is an excellent photonic platform due to its mature fabrication, low propagation loss\cite{puckett2021422}, large refractive index\cite{soref2010mid}, and wide transparency window that spans all of the identified 2D material SQEs, as shown in Fig. \ref{fig:intro}a; however, commonly used stoichiometric silicon nitride (Si$_3$N$_4$) has a strong background emission that overlaps with many SQEs\cite{elshaari2021deterministic}, degrading single-photon purity. Previous on-chip integration strategies have placed hBN and WSe$_2$ directly above or below waveguides, which has limited the coupling efficiency to only a few percent. Low coupling is exacerbated by the random position and dipole orientation of the SQEs\cite{peyskens2019integration,errando2021resonance,kim2019integrated,li2021integration}. Even for perfectly positioned and aligned emitters embedded in a single-mode waveguide, the theoretical maximum coupling efficiency is limited to $\sim 20-30\%$. 

Here, we demonstrate a novel method for efficient on-chip coupling by integrating 2D SQEs with a waveguide-coupled microring resonators using a CMOS-compatible process. Our approach is universal in that it enables the deterministic integration of SQEs with low auto-fluorescing, single-mode silicon nitride photonic circuits with precise control over the emitter placement and dipole orientation within the waveguiding structures—both of which are critical for achieving efficient emitter-mode coupling. As an example, we demonstrate an important advancement for quantum PICs by aligning an hBN SQE generating high-purity single photons at room temperature with a waveguide-coupled microring resonator (Fig. \ref{fig:intro}b). We demonstrate $46\%$ emitter-cavity coupling, which requires only a modest Purcell factor of $0.86\pm0.15$ to surpass the waveguide coupling efficiency in prior studies by nearly an order-of-magnitude\cite{kim2019integrated,elshaari2021deterministic,li2021integration}. We demonstrate the universality of the approach by coupling exciton emission from embedded WS$_2$ achieving > $63\%$ efficiency with a Purcell factor of $1.44\pm0.25$. We present various emitter-microresonator designs, coupling schemes, and metrics that provide a roadmap for the integration of SQEs spanning the UV to telecommunications wavelength regimes. Guided by a semi-classical cavity-emitter model, routes towards achieving high-purity and high-indistinguishability single-photon emission from a variety of 2D material emitters are proposed, paving the way for enabling scalable and manufacturable integrated quantum photonics with on-demand sources in silicon nitride.

\vspace{-10pt}
\section*{2D-compatible, low-emission silicon nitride}
\vspace{-5pt}

\begin{figure*}[t] \centering
     \includegraphics[scale=0.9]{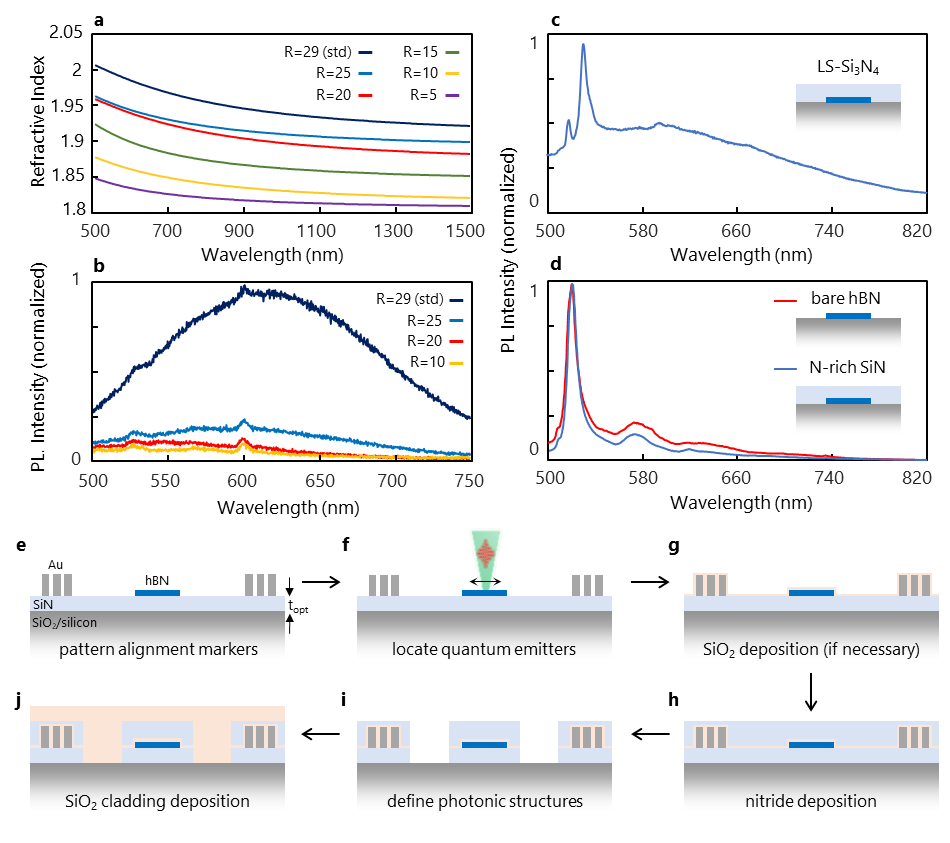}  
          \caption{\textbf{Site-specific and momentum aligned integration of hBN SQEs to low emission silicon nitride.} \textbf{a}, Refractive index of silicon nitride as function of decreasing silane/ammonia ratio, \textit{R}. \textbf{b}, Background emission of silicon nitride as function of  \textit{R}. At $R=20$, the background emission is sufficiently suppressed for high purity measurements of hBN SQEs. \textbf{c}, PL spectrum of an hBN SQE covered with 100 nm of low-stress stochiometric PECVD Si$_3$N$_4$. \textbf{d}, PL spectra of an hBN SQE before (red) and after (blue) 100-nm deposition of nitrogen-rich non-stochiometric PECVD SiN and 1000 \degree C rapid thermal anneal, acquired at the same excitation power and integration time as \textbf{c}. \textbf{e}, To position the flake with respect to the photonic structure, thin hBN flakes are exfoliated on PECVD SiN films with optional thickness ($t_{opt}$) that is either equal to full or one-half of the designed thickness of the waveguide for top or embedded flakes, respectively. Alternatively, to position the flake on bottom of the waveguide, hBN can be exfoliated directly on the SiO$_2$ substrate. Gold alignment markers are patterned with electron-beam lithography in close proximity to the flake. \textbf{f}, The position and dipole orientation of quantum emitters are determined by high numerical-aperture polarization-resolved microscopy and raster scanning the sample. \textbf{g}, Thin-film PECVD SiO$_2$ for protecting TMD flakes from damage during SiN PECVD. \textbf{h}, Deposition of remaining SiN, if necessary, to complete the photonic layer. \textbf{i}, Electron-beam lithography and ICP-RIE etching are used to define the photonic circuits. \textbf{j}, Final PECVD SiO$_2$ for the cladding layer.}   
          \label{fig:Integration}
\end{figure*}

\noindent Standard plasma-enhanced chemical vapor deposition (PECVD) of stoichiometric Si$_3$N$_4$  suffers from significant background fluorescence at visible wavelengths\cite{elshaari2021deterministic,smith2020single,senichev2021room}, which is a problem for many 2D materials that emit in this range including hBN and TMDs. To address these challenges, we extend on previous developments of nitrogen-rich silicon nitride (SiN) that eliminate the background fluorescence. Careful tuning of the PECVD RF power, voltage bias, and silane to ammonia ratio (\textit{R}) allows for the deposition of high-quality silicon nitride with negligible background fluorescence and high refractive index, without damaging the underlying 2D materials (Figure \ref{fig:Integration}). 

Similar to previous studies\cite{Austin1985,Kistner2011,senichev2021room}, we find that for decreasing $R$, the background fluorescence quenches with only a moderate reduction in the refractive index; however, creating SQEs in hBN is typically achieved \cite{vogl2018fabrication} through rapid thermal annealing up to 1100 \degree C. In previous studies, annealing has introduced or activated defects, which enhances the broad fluorescence background even in nitrogen-rich films (see supplementary material); however, by pre-conditioning the annealing chamber with an optimized oxygen/nitrogen environment, the formation of the defect band in SiN thin films annealed up to 1000 \degree C remains negligible. This points to extrinsic defects being introduced from the PECVD chamber into the nitride films as one of the primary sources of broadband fluorescence, which warrants future follow-up investigations. Up to now, the inability to anneal nitride thin films has been a severe limitation for site-controlled integration of 2D materials\cite{elshaari2021deterministic}, particularly hBN, aligned and integrated with waveguides for optimal mode coupling. 

\begin{figure*}[bth!] \centering
     \includegraphics[scale= 0.85]{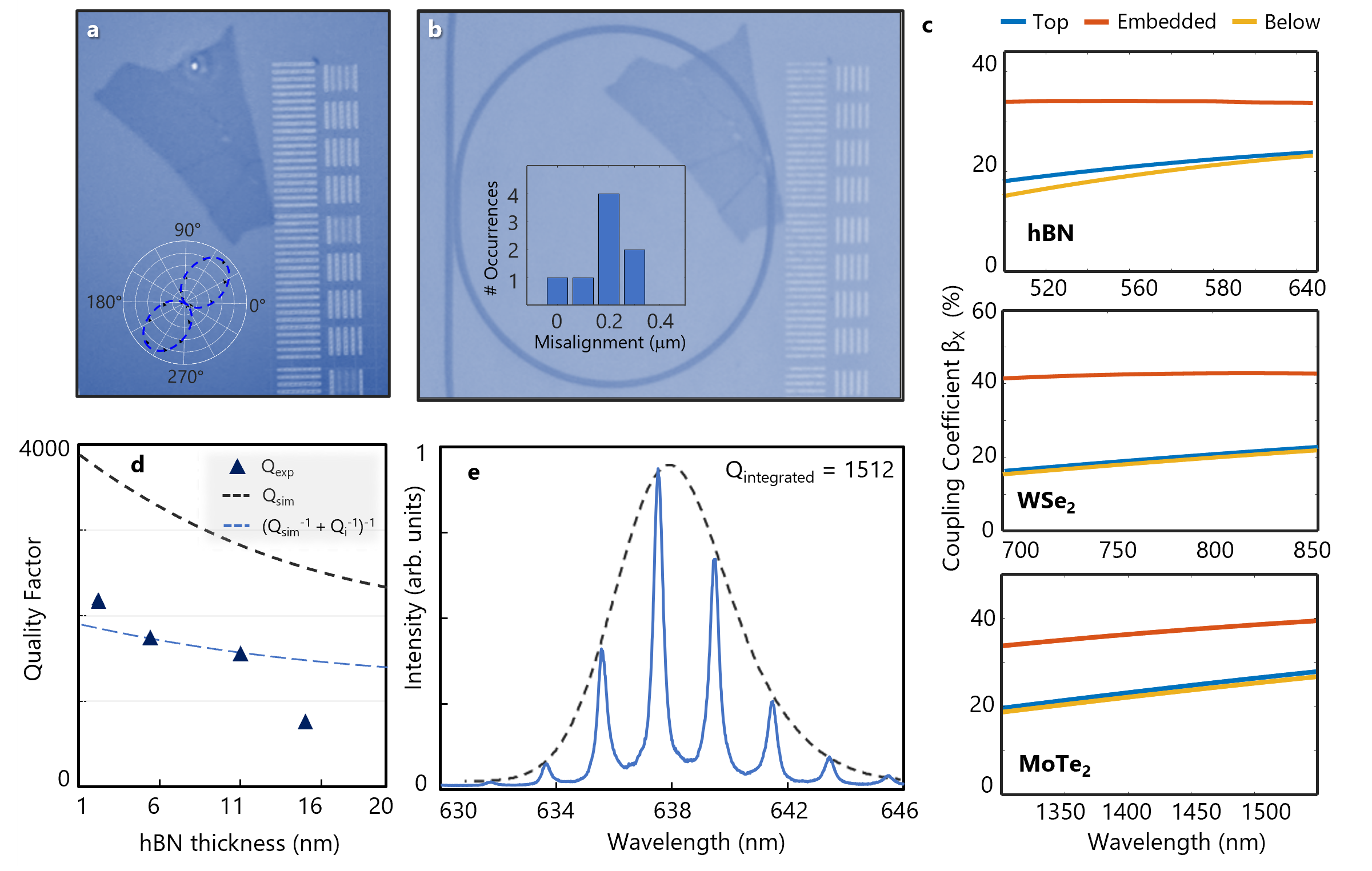}  
          \caption{\textbf{Integration of hBN SQE and microresonator characterization.} \textbf{a} Optical image of an SQE in an hBN flake near positioning markers. The SQE location is denoted by the bright white dot which corresponds to the dimmed excitation laser spot. Inset plot depicts the emission dipole orientation of the emitter. \textbf{b} Fabrication of a complete device with hBN emitter integrated and dipole-aligned to a microring resonator with 100-nm precision. Inset plot shows the distribution of the SQE positioning over several trials. \textbf{c}, Simulated waveguide-emitter coupling efficiency ($\beta$) for 2D quantum emitters on top, within, or below the straight waveguide. A maximum of up to 40$\%$ coupling efficiency (summed over both waveguide propagation directions) is possible for each type of emitter. \textbf{d}, Theoretical (lines) and experimental (points) resonator quality factor as a function of the integrated hBN flake thickness. It is assumed the flake covers $25\%$ of the ring in the simulations. Black dashed line represents the theoretical simulations, which do not take into account the intrinsic loss due to the flake and absorption in the guiding medium. Blue dashed line represents the theoretical simulations corrected with the experimental intrinsic quality factor. \textbf{e}, Characterization of a ring with integrated hBN flake. A broadband superluminescent diode (centered roughly at 638 nm, dashed line) is coupled to the output port of the waveguide. Scattered light from the ring (solid line) is collected using 0.9 NA objective. Free-spectral range (FSR) of 2.1 nm and a loaded quality factor (\textit{Q}) of 1512 is measured with an embedded flake.}
          \label{fig:microring}
\end{figure*}

Figure \ref{fig:Integration}c illustrates the results from this process. A room temperature photoluminescence spectrum is shown in Fig. \ref{fig:Integration}c from a representative hBN emitter in a flake under a 100-nm-thick low-stress stoichiometric Si$_3$N$_4$  film (see supplementary information). Nearly 50$\%$ of the emission at the zero-phonon line (ZPL) wavelength of the emitter near 540 nm arises from the broad background emission from Si$_3$N$_4$. Using the new thin-film deposition procedure, we observe pristine emission from hBN emitters with negligible background from the SiN, shown in Fig. \ref{fig:Integration}d for bare hBN (red curve) and the same hBN post growth and anneal of 100 nm of SiN (blue curve).


\vspace{-10pt}
\section*{Microresonator Design and Integration}
\vspace{-5pt}


\noindent The fabrication process for deterministically embedding 2D flakes within SiN photonic structures is illustrated in Fig. \ref{fig:Integration}(e-h) and described in more detail in the methods section and the supplementary information. A custom microscope integrated with a micro-photoluminescence (micro-PL) setup enables pre-screened SQEs to be aligned to microresonators with 100-nm precision. To do this, first gold alignment marks are patterned near the SQE flake using electron-beam lithography and lift-off techniques. The flake is screened for SQEs using polarization-resolved micro-PL to determine their position and dipole orientation with 100-nm and $\sim5\degree$ precision, respectively, using edge detection and image matrix transformation techniques. The combination of precise SQE placement and orientation with low-background, damage-free SiN deposition are the key enabling capabilities for 2D material PIC integration with single-mode photonic structures, which has proven to be difficult for randomly positioned SQEs. Since the PECVD process and annealing can be used to directly grow the SiN layer on top of the hBN SQE without altering its spectrum and single-photon purity, the flake can be integrated either on the top, in the middle, or underneath the structures.

\begin{figure*}[t!] \centering
     \includegraphics[scale=0.9]{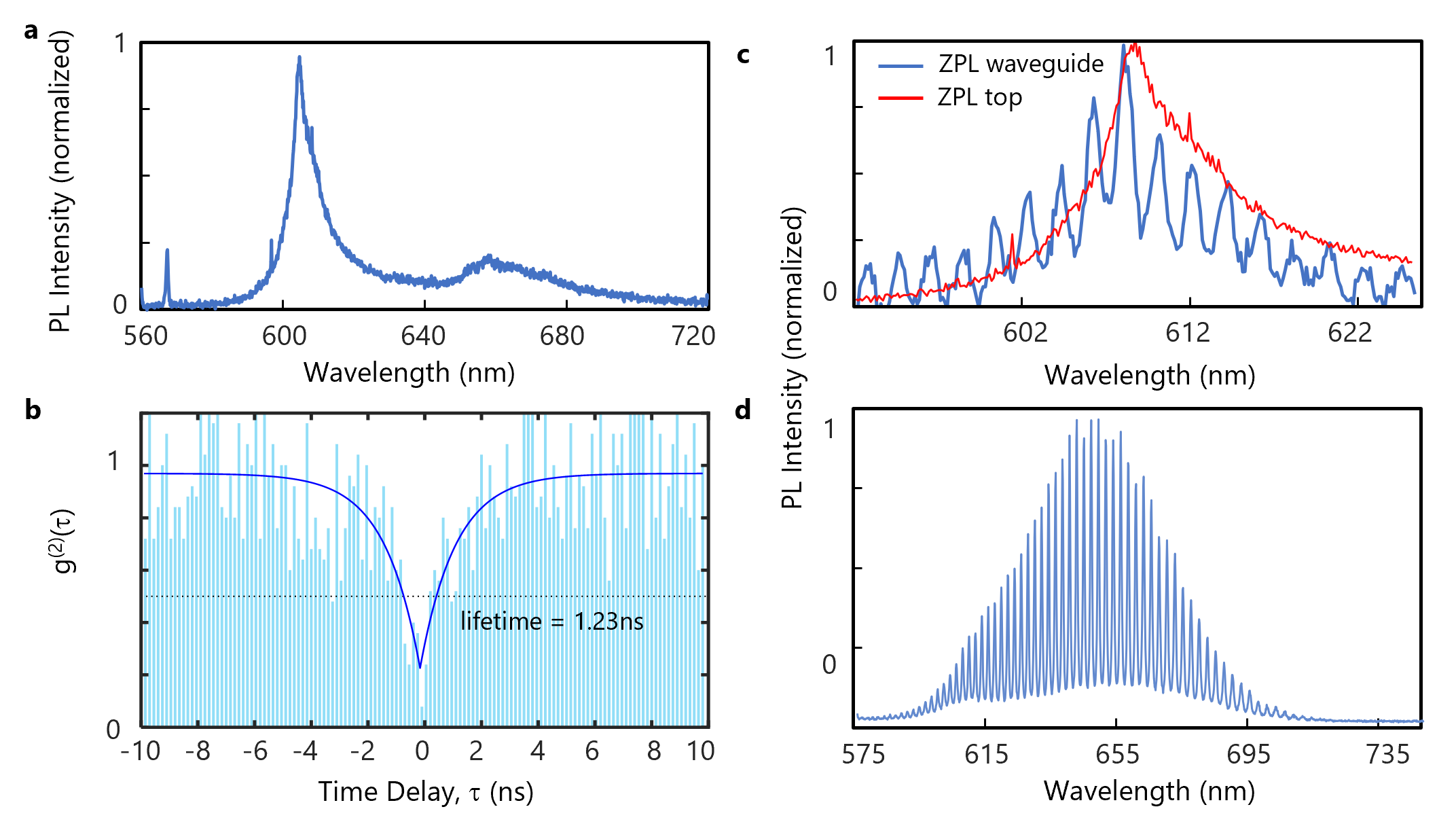}  
          \caption{\textbf{Microresonator-integrated hBN quantum emitter with high coupling and Purcell enhancement.} \textbf{a}, PL spectra of an hBN SQE acquired with top-down excitation and collection after microresonator integration. The emitter properties are preserved after the fabrication process. \textbf{b}, Second-order autocorrelation measurement demonstrating 78\% purity and a lifetime of 1.23 ns ($82\%$ purity when background-corrected). The data are raw with no correction for the background or detector dark counts. \textbf{c}, ZPL of the emitter observed from the output port of the waveguide using an aligned fiber array (blue line) and from the top collection (red line). A factor of ten reduction of the ZPL linewidth is observed (from 7.2 nm down to 0.72 nm) as expected from the bandwidth of the microresonator. The peak intensity of the ZPL is misaligned from the nearest cavity resonance by $\sim $0.35 nm. \textbf{d}, Excitonic response of a bilayer WS$_2$ collected from the waveguide port. The modes of the microresonator are clearly visible, demonstrating a quality factor up to 2400.}   
          \label{fig:results}
\end{figure*}

Figure \ref{fig:microring}c demonstrates the theoretical coupling efficiency of a dipole emitter with perfect polarization alignment integrated with a single-mode SiN waveguide at different heights for three common types of SQEs in 2D materials. The coupling efficiency of the radiated field into the optical mode, normalized to the total radiated field, is defined as $\beta$, where $\beta = 1$ corresponds to $100\%$ emission into the waveguide mode. Intuitively, the greatest mode overlap occurs when flakes are embedded in the center of the waveguide; however, care must be taken to avoid etched hBN edges within a few hundred nanometers to the emitter, which we have found can introduce edge states that alter the charge state of the defect or act as charge traps that contribute to optical dephasing and spectral diffusion. To avoid these complications, we also explored the integration of hBN underneath the waveguiding structures in which the hBN flake is not exposed to any etched surfaces. For this configuration, a theoretical coupling efficiency of $\beta = 20\%$ (Fig. \ref{fig:Integration}c) is anticipated for waveguide-integrated SQEs. In practice, the measured waveguide coupling efficiency, however, is typically limited to $\sim 1-3\%$ primarily due to poor SQE-mode overlap and dipole misalignment\cite{li2021integration,kim2019integrated}.

Alternatively, $\beta$ can be enhanced relative to waveguides by integrating the emitter within a microring resonator. For a cavity-coupled emitter, its radiative decay rate is resonantly enhanced and becomes $\Gamma = \left(1+F\right)\Gamma_0$, where $\Gamma_0$ is the radiative decay rate in the absence of the cavity and $F\Gamma_0$ is the radiative enhancement due to the cavity \cite{gould2016efficient,faraon2011resonant}. This enhancement can be quantified through the Purcell factor $F= (3/4\pi^2)(\lambda_{zpl}/n_{cav})^3(Q/V)$ where $n_{cav}$, $Q$, and $V$ are the refractive index, quality factor, and mode-volume of the cavity. It is important to point out that for an emitter weakly coupled to the cavity, it is often assumed that the density of states of the free-space modes remains undisturbed\cite{faraon2011resonant,fox2006quantum}. As a result, $\beta$ for a cavity-coupled emitter can be expressed as $\beta = F/\left(1+F\right)$. This has important implications for exploiting resonant cavity effects: increasing $F$, and thus $\beta$, amounts to designing high $Q$ cavities with small mode volume. As we show below, even for $F \sim 1$, the SQE emission into a photonic circuit can be significantly enhanced relative to an emitter coupled to a waveguide, motivating the need for on-chip cavity enhancement.

The Purcell factor is typically defined in the "good-emitter" regime in which the cavity linewidth $\kappa$ is larger than the SQE linewidth $\gamma$; however, in many instances, including hBN emitters at room temperature, phonon-induced dephasing broadens the ZPL width beyond the radiative limit, and the cavity-coupled system is found in the so-called “bad-emitter” regime, i.e. $\gamma > \kappa$, where only a portion of the SQE ZPL couples into the cavity. This effectively reduces the traditional Purcell factor to $F \kappa/ \gamma$, where $\kappa/ \gamma$ heuristically represents the ratio of the radiated power from the SQE that overlaps with the cavity mode. In the bad emitter regime, a wavelength dependent Purcell factor and coupling efficiency can be defined as a function of the spectral power of the emitter and are given by $F_s(\lambda) = I_{cav}(\lambda)/I_{fb}(\lambda)$ and $\beta_s= F_s(\lambda)/\left(1+F_s(\lambda)\right)$  where $I_{cav}(\lambda)$ and $I_{fb}(\lambda)$ are the spectral intensity of the emitter into the cavity mode and into free-space, respectively. This, in effect, negates the $\kappa/ \gamma$ factor and allows for the enhancement of the portion of the ZPL that is resonantly coupled to the cavity to be quantified regardless of the emitter-cavity regime\cite{kaupp2013scaling}. Whether in the good- or bad-emitter regime, $\beta_s\left(\lambda_0\right)$ specifies the cavity enhancement at the emission wavelength $\lambda_0$, while integration over $\lambda$ determines the total $\beta$. For waveguide-coupled emitters without any resonant effects, $\beta_s\left(\lambda\right) = \beta$.

As shown below, to maximize the total system efficiency, $Q$ should be chosen to be in an intermediate region between the good- and bad-emitter regimes. Hence, we chose a racetrack resonator configuration with a 100-nm-thick and 600-nm-wide cross section, a 3-$\mu$m-long coupler region that results in a free spectral range (FSR) of 2 nm, a mode-volume of $30(\frac{\lambda}{n})^3$ at the resonance of interest around 610 nm, and a simulated quality factor of 7000, which is comparable to the cryogenic linewidth of hBN emitters observed in our samples. We first characterized resonators without embedded 2D materials to establish a baseline for our quality factor, which we can write as $Q^{-1} = Q_i^{-1} + Q_c^{-1} + Q_{sc}^{-1}$, where $Q_i$ corresponds to the bare resonator, $Q_c$ corresponds to the coupling to the waveguide, and $Q_{sc}$ arises from additional scattering from an integrated 2D flake. Measurements from 10 nominally identical resonators from three different fabrication runs yield an average $Q_i= 3560$ and $Q_c = 9700$, which results in a slightly lower $Q$ than our simulations likely due to etched sidewall roughness and a larger waveguide-resonator coupling gap.

We next examined the impact of integrated hBN flakes on the resonator $Q$. As expected, with increasing flake thickness, $Q$ decreases, which matches our simulations (Fig. \ref{fig:microring}d). While hBN has a similar refractive index as SiN, light scattering at the SiN-hBN interfaces, which has a more pronounced effect for thicker flakes, dominates the loss and reduction of $Q$ in our simulations. Experimentally, for flakes with a thickness exceeding 30 nm, $Q$ decreases by an order-of-magnitude. Given that hBN emitters tend to have narrower linewidth and brighter emission in thin, but multilayer flakes, this result confirms an important design trade-off for resonator integration. We found that emitters with linewidths as narrow as 3-4 nm at room temperature are routinely identified in $\sim 15$-nm-thick flakes. Figure \ref{fig:microring}e shows the spectrum of the microresonator with an hBN flake integrated below the ring, exhibiting a loaded $Q = 1512$. We note that while we focus on these emitters in this study as a demonstration, the integration strategy and fabrication procedure are entirely 2D materials agnostic. For example, as we later show, integration of 2D TMDs, unlike hBN, does not introduce significant scattering loss in the cavity and the $Q$ remains close to the empty microring resonator.

\vspace{-10pt}
\section*{Cavity-Enhanced 2D Material On-Chip Coupling}
\vspace{-5pt}

\noindent Figure \ref{fig:results}a shows the photoluminescence (PL) spectrum of a representative hBN emitter after the complete device fabrication using top-down excitation and collection at room temperature. As demonstrated in Fig. \ref{fig:Integration}d and Fig. \ref{fig:results}a, the integration and fabrication do not degrade the quality of the emitter and we retain the background-free emission. The photon anti-bunching behavior verifying single-photon emission is illustrated in Fig. \ref{fig:results}b in which $g^{\left(2\right)}\left(0\right) = 0.22$ (0.18 with background correction). We next excite the emitter using a 0.9 NA objective from above the resonator, and emission into the waveguide is collected into a single-mode fiber and sent to a spectrometer and charge-coupled-device camera. Figure \ref{fig:results}c shows the integrated hBN SQE with the ZPL emission centered near 610 nm. The dashed line is emission collected from above the emitter, whereas the solid line is emission collected from the waveguide. Bright emission into the microresonator modes is clearly observed by the peaks imprinted on the ZPL emission spectrum separated by an FSR of $\sim$ 2 nm. A similar response is imprinted on the room temperature exciton emission from monolayer WS$_2$ in Fig. \ref{fig:results}d, demonstrating the universality of the approach for 2D material integration.

To extract the spectral Purcell enhancement $F_s(\lambda)$ and the spectral coupling efficiency $\beta_s(\lambda)$, we follow a procedure previously reported for SQEs \cite{kim2019integrated,li2021integration,gould2016efficient}. Accounting for the optical loss in our system, the effective Purcell enhancement at the ZPL peak wavelength can be expressed as

\begin{equation}
F_s = \frac{\eta_{ob}  \eta_{top} }{\eta_{out}  \eta_{facet}  \eta_{side} }\frac{I_{cav}^{ccd}}{I_{fb}^{ccd}},
\end{equation}

\noindent where $\eta_{ob}$, $\eta_{top}$, $\eta_{out}$, $\eta_{facet}$, $\eta_{side}$, $I_{fb}^{ccd}$, and $I_{cav}^{ccd}$ are respectively the portion of the total free-space light collected by the top objective, efficiency of the top collection path, microring out-coupling efficiency, facet coupling efficiency, side path collection efficiency, spectral intensity measured on the CCD at the ZPL wavelength from the top objective, and spectral intensity measured from the waveguide output port (see supplementary information). 

\begin{figure}[b!] \centering
     \includegraphics[scale=0.8]{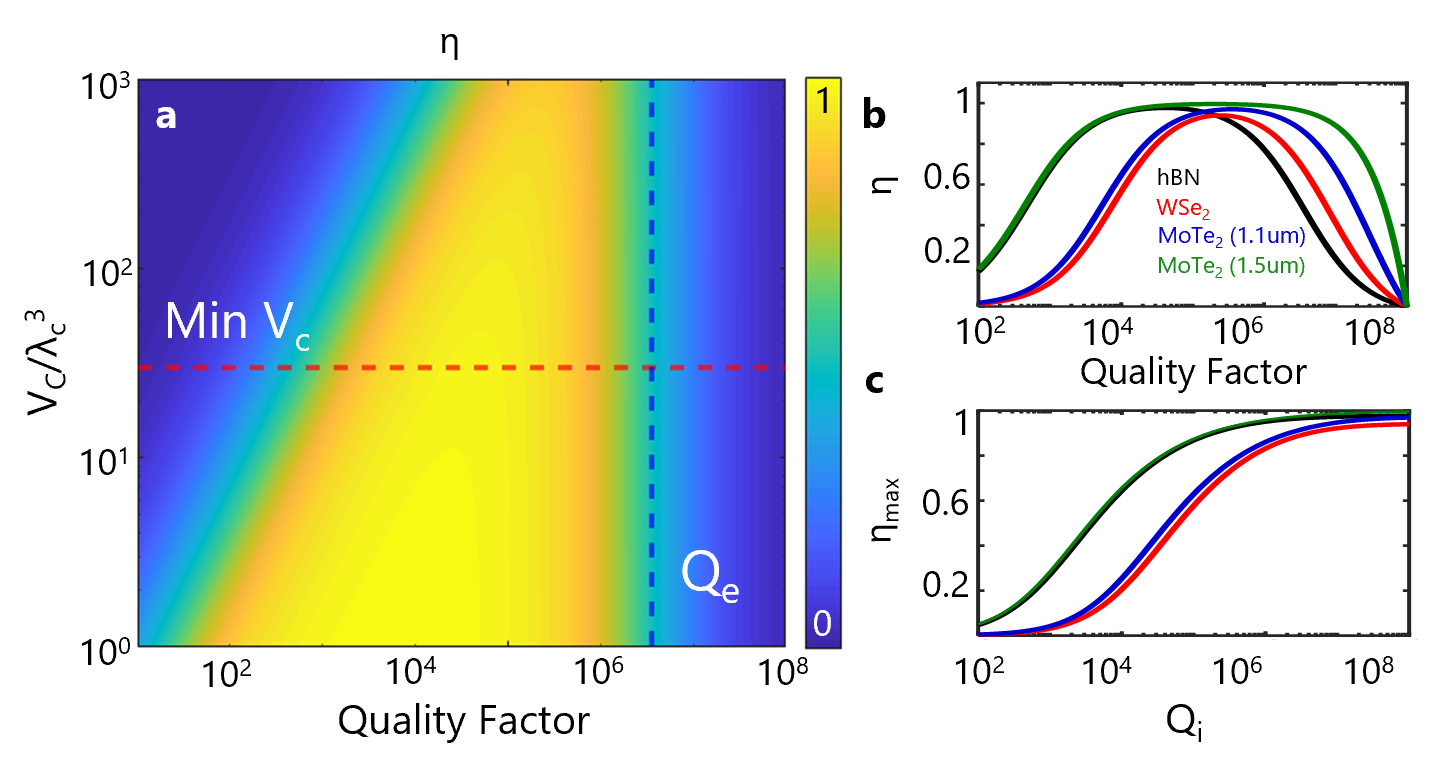}  
          \caption{\textbf{Projected performance of 2D material quantum emitters in a SiN heterogeneous platform }. \textbf{a}, Mode volume versus coupling quality factor $Q_c$ for hBN visible emitters. The red dashed line demonstrates the mode volume achieved in our SiN microresonators. The blue dashed line demonstrates the intrinsic quality factor $Q_e$ of the emitter determined from its linewidth. \textbf{b}, Total system efficiency $\eta$ as function of loaded $Q$ at the minimum achievable mode volume for hBN, WSe$_2$, and MoTe$_2$ emitters in the visible, near infrared, and telecom wavelength. The radiative lifetime of hBN, WSe$_2$, MoTe$_2$ (1.1 $\mu$m), and MoTe$_2$ (1.5 $\mu$m) is taken from previous studies to be 1.2 ns, 4 ns, 22.2 ns, and 1 $\mu$s with quantum efficiencies of 0.87, 0.05, 0.07, and 0.86, respectively\cite{nikolay2019direct,parto2021defect,zhao2021site}. \textbf{c}, Maximum total system efficiency $\eta_{max}$ achievable for each class of emitters as function of the intrinsic quality factor of the SiN platform.}   
          \label{fig:cavityoptimize}
\end{figure}

For the integrated emitter in Fig. \ref{fig:results}c, a spectral Purcell factor up to $0.86\pm0.15$ corresponding to $\beta_s = 46\pm4\%$ is measured on the ZPL resonance ($\beta = 28\pm4\%$ integrated across the entire spectrum). The deviation from the theoretical estimate of the effective Purcell factor for this system (equal to 1.7) can be attributed to small misalignment and dipole orientation inaccuracies. Importantly, even though $F_s$ is close to unity, this results in nearly half of the emission now coupled into the cavity mode. This is best reflected in $\beta_s$ of the cavity-coupled system. Here, even the lower-bound of our measured $\beta_s$ exceeds the maximum theoretical coupling efficiency into a waveguide of the same configuration ($\sim 20\%$ as shown in Fig. \ref{fig:Integration}). Similarly for the integrated WS$_2$, $\beta_s= 63\pm4\%$ is obtained, amounting to a spectral Purcell factor of $F_s = 1.44\pm0.25$. The higher measured Purcell factor for WS$_2$ is due to the fact that it is thinner, and thus it does not significantly alter the loaded quality factor of the resonator.

\section*{\label{sec:Discussion}Discussion}
\noindent

\noindent To design a 2D quantum emitter-cavity system, it is important to consider the total PIC efficiency, which can be expressed as $\eta = \eta_{qe} * \beta *\eta_{out}$, where $\eta_{qe}$ is the quantum efficiency of the emitter and $\eta_{out}$ is the extraction efficiency of the coupled light in the cavity into the bus waveguide\cite{thomas2014waveguide,gould2016efficient}. Maximizing $\eta$ is a multi-variable problem because the individual component efficiencies cannot be adjusted independently. For instance, one can reach the over-coupled regime using a small waveguide-resonator gap or a pulley coupled bus waveguide, which leads to $\eta_{out} \rightarrow 1$ as $Q_c$ decreases; however, a significant reduction of $Q_c$ lowers the enhancement and the $\beta$ factor. Similarly, while a high $Q$ results in a larger $\beta$, for large $Q$, the cavity-emitter system can enter the bad-emitter regime where only a portion of the ZPL will couple into the microring resonator and $\eta$ will begin to decrease. Generally, the linewidth of the emitter sets a practical upper bound for the loaded $Q$. While this can imply that SQEs with the narrowest linewidth are more suitable for cavity integration, it should be noted that the SQE internal quantum efficiency $\eta_{qe}$ plays an important role in the emitter-cavity design. For instance, at cryogenic temperatures, while WSe$_2$ SQEs have a relatively narrow linewidth ($\sim 100\ \mu$eV) compared to hBN emitters, $\eta_{qe}$ for WSe$_2$ has been estimated\cite{luo2019single,parto2021defect} to be only $\sim1-5\%$ compared to up to $87\%$ reported for hBN\cite{nikolay2019direct}. Therefore, to optimize the cavity design with high system-wide efficiency for different 2D emitters, a holistic approach must be considered. 

In Fig. \ref{fig:cavityoptimize}, we explore the performance of a 2D SQE-cavity system using solutions to the Jaynes-Cummings Hamiltonian for a two-level system interacting with a cavity (see supplementary information). Initially, we set $Q_i$ to $10^7$ to explore the regime in which the coupling $Q_c$ governs $Q$ and its effect on $\eta$. Figure \ref{fig:cavityoptimize}a shows $\eta$ as function of mode volume and the microresonator quality factor $Q$ for hBN emitters at cryogenic temperatures, where it assumed that pure dephasing from phonons is negligible and the linewidth is governed by recombination. The vertical dashed line sets the boundary of the bad-emitter regime in which the total quality factor exceeds the emitter quality factor $Q_e$ determined from its linewidth; as $Q_c$ becomes larger than $Q_e$, only a portion of the emitter couples in to the cavity. The horizontal dashed line in Fig. \ref{fig:cavityoptimize}a indicates the minimum mode volume for our PICs ($30(\lambda/n)^3$), and slices along this line are shown in Fig. \ref{fig:cavityoptimize}b for hBN and TMD emitters. For $Q \sim 10^4$, $\eta$ up to $95\%$ is possible for hBN emitters at cryogenic temperatures. For $Q$ exceeding $Q_e$, the system enters the bad-emitter regime and the extraction efficiency decreases. 

Purcell enhancement can compensate for intrinsic low quantum efficiency of some emitter types, such as WSe$_2$ and MoTe$_2$ ($1.1 \mu$m emission), provided loaded $Q > 10^5$ is reached for the platform. On the other hand, integration of emitters with high quantum efficiency, such as hBN and MoTe$_2$ (1.5 $\mu$m emission) can be realized at lower $Q$. At the moment, the primary limitation for achieving unity extraction efficiency for ideal 2D emitters is the intrinsic quality factor of the platform, which sets an upper bound on the achievable $Q$. This is shown in Fig. \ref{fig:cavityoptimize}c, which represents the maximum attainable extraction efficiency for each class of the emitters as function of the intrinsic quality factor of the platform. As the intrinsic $Q$ approaches $10^6$, which is already achievable for different SiN waveguide aspect ratios, it is possible to integrate 2D quantum emitters with total efficiencies exceeding $80\%$.


Taken together, our simulations and experiments provide a straightforward approach for deterministically aligning and orienting SQEs in 2D materials to on-chip microresonators with a route towards high coupling efficiency. Already this approach achieves $46\%$ coupling efficiency into the resonator at the emitter ZPL resonance, which is an order of magnitude higher than waveguide coupling for hBN. System-wide efficiency approaching $10\%$ can be attained in the near-term with modest improvements to the microresonator $Q$ and its design for over-coupling. The platform and methods developed in this work can serve as a stepping stone towards future demonstrations of on-chip 2D quantum emitter integration with high extraction efficiency, brightness, and indistinguishably. In the near future, by exploring SiN microresonators embedded with other SQEs with narrow linewidths, such as WSe$_2$ and MoTe$_2$, new opportunities exist for on-demand and site-controlled SQEs with silicon-based photonics for chip-scale quantum information applications.

\begin{acknowledgments}
\noindent This work was supported by NSF Award No. ECCS-2032272 and the NSF Quantum Foundry through Q-AMASE-i program Award No. DMR-1906325. Experiments were performed with support from DURIP Award No. FA9550-21-1-0257. S.I.A. acknowledges support from the California NanoSystems Institute through the Elings fellowship. K.W. and T.T. acknowledge support from JSPS KAKENHI (Grant Numbers 19H05790, 20H00354, and 21H05233). K.P. and S.A. contributed equally to this work.
\end{acknowledgments}

\section*{Data Availability Statement}

\noindent The data that support the findings in this study are available from the corresponding author upon reasonable request.

\section*{References}

\bibliography{references}

\section*{Methods}

\subsection*{2D SQE compatible silicon nitride PECVD growth and anneal}

\noindent The background-free integrated photonic platform was realized by growing 100 nm nitrogen-rich SiN films on 3 $\mu$m thick thermal oxide substrates. All films were grown using a vision 310 advanced vacuum PECVD system at 300 \degree C and 800 mTorr with iterative 30 W, 13.56 MHz and 110 W, 187 kHz plasma steps for 8 and 1.5 seconds, respectively. The silane, ammonia, and nitrogen precursor flow rates were chosen to be 360 sccm, 18 sccm, and 980 sccm, respectively, to achieve nitrogen-rich growth conditions that quenches the SiN background emission while maintaining a high refractive index. A 20-minute, 1000 \degree C rapid thermal anneal was performed to create optically active defects in hBN that was first transferred onto the Si$_3$N$_4$. Annealing was carried out in a 100 sccm nitrogen and 100 sccm oxygen environment to create bright room temperature stable emitters in hBN without activating the SiN defect band.\\

\noindent Figure \ref{fig:Integration}a shows the measured refractive index of SiN for $R = 29$ (stoichiometric) to $R = 5$. With decreasing $R$, the refractive index decreases from 2.01 to 1.86 at 500 nm. We find that at $R=25$, the background fluorescence begins to quench, as shown in Fig. \ref{fig:Integration}b. Because of the trade-off between the reduction of the refractive index and the fluorescence quenching with decreasing $R$, we find that a ratio of $R=20$ provides sufficient background-free SiN with only a moderate reduction of the index. By fine-tuning the PECVD parameters, it is possible to simultaneously achieve quenching while maintaining a refractive index > 1.9 without degrading the optical quality of the embedded 2D materials.\\

\vspace{-10pt}
\subsection*{Numerical Simulations}
\noindent All simulations have been performed using Ansys Lumerical. Emitter-waveguide coupling efficiency has been calculated using Lumerical FDTD using a dipole source as an emitter. Coupling factor to the fundamental mode is calculated using a mode expansion monitor. Ring quality factor and free spectral range are simulated  using Lumerical varFDTD with the fundamental TE mode source as an excitation. The mode volume of the cavity mode is calculated with FDTD using the formula $V = \frac{\int \epsilon E^2 \,dv}{max(\epsilon E^2 )}$. Top collection efficiency of the photoluminescence using high NA objective (0.95) are performed using Lumerical FDTD through the integration over the acceptance cone. More details on the simulations can be found in the supplementary information.

\subsection*{Photonic Device Fabrication}

\noindent The fabrication process for integrating 2D quantum emitters within SiN photonic structures is illustrated in Fig. \ref{fig:Integration}e-\ref{fig:Integration}j and is described in more detail in the supplementary information. Briefly, SiN thin films were grown by PECVD on a 3-$\mu$m-thick SiO$_2$ film on silicon, where the film thickness is determined by the desired position of the flake in the waveguide structure normal to the sample plane. A variety of SiN films were grown with an increasing ammonia-to-silane ratio to characterize the 2D material transfer, annealing, and full fabrication process. Using an all-dry visco-elastic transfer technique, 2D material flakes of hBN or WS$_2$ are transferred within the electron-beam processing window of 200 $\times$ 200 $\mu$m. For hBN flakes, a 5-minute, 250 watt O$_2$ plasma and rapid thermal annealing step at 1000 \degree C activates the quantum emitters, which are then pre-screened by scanning the sample in a 0.9 numerical-aperture (NA) photoluminescence microscopy setup. The emission brightness and single-photon purity are the primary metrics we use to identify suitable emitters for integration. Once the target emitters and flakes are identified, electron-beam lithography and lift-off techniques are used to pattern metal alignment markers and a grid for subsequent electron-beam lithography steps. The grid allows for identification of emitters with $\sim$100-nm precision, which allows for post-fabrication of structures carefully aligned to the emitters. For transition metal dichalcogenides, we next deposit a thin ($\sim$ 10 nm) film of PECVD SiO$_2$ as a protection layer, which our simulations show has negligible impact on the confined optical modes, propagation loss, and resonator quality factor. Next, the remaining SiN thin film is grown, followed by an electron-beam lithography and reactive-ion etching step, and finally an optional $0.1 \sim 1$ $\mu$m PECVD SiO$_2$ cladding layer. Note that the cladding layer can lead to decrease in the coupling efficiency for bottom hBN configurations as illustrated in Fig. \ref{fig:Integration}c. However, the use of cladding increases the efficiency of the edge couplers. We use inverse taper couplers to expand the mode to match with single-mode fiber.

\subsection*{Single-Quantum Emitter Integration}

\noindent To align the hBN SQEs to the resonators, first, a fine alignment metal bar array structure (500-nm spaced bars) is patterned in a location approximately 20 $\mu$m away from the SQE using electron-beam lithography and standard lift-off techniques. Next, using our customized blue light microscope (infinity corrected objective with 0.9 NA, 1 mm working-distance objective coupled with a 50-mm focal length achromatic doublet, and zoom lens), optical images of the flakes are taken when the laser spot is centered at the location of the SQE (determined via maximum photoluminescence intensity) and with the grid pattern in the 70 $\mu$m by 70 $\mu$m field of view (FOV). Using an edge detection code based on the "canny edge detection" technique, the edges of the fine-alignment grid are detected. Then the image scale undergoes matrix transformation until the mean square error of the difference between the actual and the optically detected patterns are minimized. Afterward, the center of the laser spot is considered as the SQE's location, and the ring pattern is aligned to the defect based on the measured emission dipole orientation of the defect. Note that the initial spacing of the fine-alignment metal bars was designed so that at least 2 $\mu$m spacing would exist between the alignment marks and the microring, ensuring no loss due to metal. The key step of this process is using both horizontal and vertical alignment bar grids that are visible in the FOV of the image, which allows for a sufficient amount of identified edges along both directions that can be used to correct any image deformations. The emitter-cavity alignment accuracy of the process, determined over nine different trials, is approximately 100 nm. This can be reduced in the future by using closed-loop piezo stages or scanners. 

While we focus on optimizing $\beta$ in this study, it is also equally important to maximize $\eta_{out}$, which can be expressed as the ratio $Q/Q_c$ with $Q$ and $Q_c$ defined as the loaded and coupling quality factors of the cavity. For the cavity-coupled system shown in Fig. \ref{fig:results}c, $\eta_{out}$ is estimated to be $7\%$ of the total coupled light into the cavity, leading to total system efficiency of $2.8\%$. This is a direct result of the resonator being under-coupled ($Q_c > Q_i$). By extending the waveguide-cavity coupler, $Q_c$ can be decreased to $\sim$3000 with only a $14\%$ reduction of the loaded quality factor and the Purcell factor. This would result in a significant enhancement of the cavity out-coupling efficiency to $\eta_{out} > 17\%$ and the total system efficiency to $7\%$. \\

\vspace{-20pt}

\subsection*{Optical characterization}

For the room-temperature steady-state photoluminescence spectroscopy, a custom-built infinity-corrected microscope assembly equipped with 0.9 NA, 1 mm working distance dry objective was used. A single-mode green diode laser emitting at 532 nm was used for non-resonant excitation of the emitters at various powers up to 100 $\mu$W. Two dichroic mirrors centered at 510 nm and 540 nm were used to separate the white light imaging path and excitation/collection path while maximizing the collection efficiency. A 600 nm long-pass spectral filter was used to extinguish the excitation laser in the collection path. Second-order autocorrelation measurements with continuous-wave excitation were performed by passing the collected emission through the side-slit of the spectrometer (Princeton instruments HRS-500) with grating (300g/mm) set at the center ZPL wavelength. 

Filtered light was collected into a multimode fiber beamsplitter and sent into two single-photon avalanche detectors (Excelitas SPCM-AQRH-13-FC) through two optical circulators to minimize cross-talk between the detectors. The electrical signals from detectors (350 ps timing resolution) were analyzed using the Swabian Time Tagger Ultra photon counting module (8 ps timing resolution). For side collection out of the waveguide, a custom-made v-groove array (4x PM460-HP from OZ Optics) mounted on a Thorlabs NanoMax stage (MAX313D) was aligned with the exposed edge-coupler facets. Output light was collimated using a 0.22 NA collimation package and was reintroduced to the collection path via a flip mirror. To characterize the microring resonator transmission spectrum, a superluminescent diode (Thorlabs SLD635T) was coupled into the single-mode v-groove PM460-HP fiber. The light was then injected into the top bus waveguide port, and transmitted light was coupled into the bottom port. 
 
\end{document}